\documentclass[12pt,a4paper]{article}
\usepackage{graphicx,amsmath,amsfonts}

\usepackage{url}

\def \ba{\begin{eqnarray}}\def\ea{\end{eqnarray}}
\def\bc{\begin{center}}\def\ec{\end{center}}
\def\nn{\nonumber\\}
\title{\huge \bf Photoproduction of   $\omega$ mesons  off   nuclei and  impact of   polarization on meson-nucleon  interaction}
\author{E. Chudakov, S. Gevorkyan, A. Somov}
\date{}

\begin{document}
\maketitle 
 \bc{\bf Abstract}\ec
 We consider photoproduction of $\omega$ mesons off complex nuclei to study interactions of transversely and longitudinally polarized 
vector mesons with nucleons. Whereas the total cross section for interactions of the transversely polarized vector mesons with nucleons 
$\sigma_T=\sigma(V_TN)$ can be obtained from coherent photoproduction, measurements  of vector meson photoproduction in the incoherent 
region provide a unique opportunity to extract the not-yet-measured  total cross section for longitudinally polarized mesons  
$\sigma_L=\sigma(V_LN)$. The predictions for the latter strongly depend on the theoretical approaches. This work is stimulated by the construction 
of the new experiment GlueX at Jefferson Lab, designed to study the photoproduction of mesons in a large beam energy range up to 12 GeV.

\section{Introduction} 
Production of unstable particles off nuclei allows one to  determine  the total cross section of the unstable particle interaction with nucleons. 
For particles  with nonzero spin,  interactions with nucleons are  defined  by a set of amplitudes corresponding to the different particle polarizations.
 
The first indication that interaction of a vector meson V ($\rho, \omega, \varphi$) with  a nucleon  can depend  on the  meson polarization comes from  the  
$\rho$   electroproduction  data. The ratio of production cross sections for a proton target can be represented as 
$R= \sigma(\gamma_L p\to V_Lp)/\sigma(\gamma_Tp\to V_Tp)=\xi\frac{Q^2}{m_{\rho}^2}$,
where   the parameter $\xi$ corresponds to the ratio of the longitudinal to  transverse $\rho^0$ total cross sections  $\xi=\sigma_L(\rho p)/\sigma_T(\rho p)$.
The value of  $\xi$ obtained from measurements is $\xi\approx 0.5$~\cite{yennie78}, while the naive quark model predicts equal cross sections 
$\sigma_L(\rho p)=\sigma_T(\rho p)$. 

The calculations of  the valence quark distribution in hadrons based on the generalized QCD sum rules  technique~\cite{ioffe00,ioffe02} suggests that the 
meson polarization can impact interactions with nucleons. In this approach  the valence quarks  distribution in transversely and longitudinally polarized 
vector mesons are significantly different. In the longitudinally polarized vector meson 80\% of the momentum is carried by valence quarks, 
leaving only about 20\%  of the momentum for gluons and sea quarks, which are responsible for strong interactionss~\cite{ioffe02}. This results in a possible dependence of 
interactions on meson polarization. According to recent calculations~\cite{forshaw10,forshaw12}, light-cone wave functions of the $\rho$ meson computed in the AdS/QCD
approach ~\cite{brodsky09} have significantly different dependence  on the light cone momentum for longitudinal and transverse polarizations, which would 
lead to different interactions of polarized mesons with nucleons ~\cite{gevorkyan15}.

Dependence of  vector particle interactions on the particle's polarization has been known for many years in the case when the constituents of the particle are  in 
the D-wave state. A good example of such effect is the deuteron interaction with matter~\cite{glauber69}. The D-wave component in the deuteron wave function leads 
to different absorption in matter for transversely and longitudinally polarized deuterons. This effect was  experimentally measured and described in ~\cite{azhg10}. 
There are also predictions that the interaction of mesons with nonzero orbital momentum with nucleons is strongly correlated with the meson polarization ~\cite{strikman98,huefner00}.
For the  ground-state S-wave vector mesons ($\rho,\omega,\varphi$) the  D-wave component in their wave functions can emerge as a result of the Lorentz transformation ~\cite{jaus91}.

The only attempt  to study the impact of the vector meson polarization on its  absorption was made many years ago~\cite{leksin78} using the charge exchange reaction 
$\pi^- + A\to\rho^0+A^\prime$. At first glance the experimental data supported the assumption that $\sigma_T(\rho N)=\sigma_L(\rho N)$. However, there are reasons against 
such a conclusion. It was  shown~\cite{tarasov75} that due to a low energy of the primary beam ($E_{\pi} = 3.7\;{\rm  GeV}$) and the large decay width of the 
$\rho$ meson some mesons decay inside the nucleus, which complicated the  interpretation of the experimental data. 

\section{ Photoproduction of $\omega$ mesons}

In the early  70's many  experiments were carried out to study  photoproduction of vector mesons on nuclei~\cite{yennie78}. These experiments had two main goals:
\begin{itemize} 
\item Extraction of vector meson-nucleon  total cross sections  $\sigma(VN)$  in order to check  quark model predictions.
\item Verification of the  vector dominance model (VDM) and finding the limits of its  validity.
\end{itemize}

The first $\omega$ photoproduction experiments at high energies using a  large set of  nuclei  were carried out by the Rochester group at Cornell~\cite{behrend70,abramson78} and 
the Bonn-Pisa group at DESY~\cite{braccini70}. The mean photon energies used at Cornell were 6.8 GeV and 9.0 GeV. The $\omega$ mesons were detected via the $3\pi$ decay mode.
The Bonn-Pisa group used  beam photons with a mean energy of 5.7 GeV. The $\omega$ mesons were reconstructed using the $\pi^0\gamma$ decay mode. Both experiments confirmed the 
quark model prediction: 
$\sigma(\omega N)=\sigma(\rho N)=(\sigma(\pi^+N)+\sigma(\pi^-N))/2$,
but the measured value of the photon-omega coupling constant $\frac{\gamma^2_{\omega}}{4\pi}$ was much 
higher than the storage ring results and SU(3) predictions~\cite{yennie78}. A later experiment at a mean energy of 3.9 GeV  was performed  at the electron synchrotron 
NINA at Daresburry~\cite{brodbeck78}. The nuclear absorption of $\omega$ mesons was found to be in agreement with the previous experiments. The value 
of $\frac{\gamma^2_{\omega}}{4\pi}$  extracted from  the experimental data was consistent with the predictions of  SU(3) although with large experimental errors. 
The discrepancies in the measured value of the  coupling constant were still not resolved.  The total cross section  $\sigma(\omega N)$ 
was extracted from the coherent part of the photoproduction cross section, whereas the incoherent part was considered  to be a background. 

Recently, the photoproduction of omega mesons was  measured at  ELSA~\cite{elsa05}  via  the  decay $\omega\to\pi^0\gamma$ and at Jefferson Lab~\cite{jlab10} using  
the rare electromagnetic decay $\omega\to e^+e^-$. The main goal was to investigate the impact of the nuclear environment on the vector mesons mass, decay width, and 
meson absorbtion. In order to have a significant fraction of the mesons decay in the nuclei, the energies of the  detected mesons were restricted to 1-2 GeV, 
where the large contribution  from nucleon resonances complicates the interpretation of experimental  results~\cite{mosel10}. Disagreements between the experimental 
measurements are not fully understood.

In all these experiments no attempt was  done to separate  absorption of transversely and longitudinally polarized omega mesons. This effect can potentially be studied 
using the GlueX detector in Hall D~\cite{gluex}, the new experimental facility constructed at Jefferson Lab. The Hall D facility provides a photon beam produced by 
12 GeV electrons using the coherent or incoherent bremsstrahlung process. The experiment will allow to study photoproduction of mesons by reconstructing both neutral 
and charged final states in a large beam energy range up to 12 GeV. Photoproduction of  $\omega$  mesons on nuclear targets in the GlueX kinematic region  is a unique 
way to study the dependence of the strong interaction on the polarization of vector mesons. The reasons are as follows:
\begin{itemize}
\item Photoproduction of $\omega$ mesons on nucleons  $\gamma N\to \omega N$ at the photon energies of several GeV is determined 
by  t-channel Pomeron exchange (diffraction, natural parity exchange ) and  one-pion-exchange (unnatural parity exchange). 
The pion exchange leads to production of longitudinally  polarized $\omega$ mesons, unlike the diffraction process, which results in the 
production of  transversely polarized mesons due to s-channel helicity conservation. The contributions from the diffraction and 
pion exchange are almost equal at a photon energy of  $E_{\gamma}=5\;{\rm GeV}$~\cite{yennie78}. Measuring the $\omega$ meson production at
different energies would provide samples with different contributions of the longitudinally polarized $\omega$ mesons.
\item  In the coherent photoproduction  $\gamma A\to \omega A$  (the nucleus left in its ground state) the unnatural exchange part of the elementary amplitude  cancels out 
since in the coherent processes the amplitudes for interactions with protons and neutrons are added with the opposite signs.
Therefore, from the coherent photoproduction one can extract only  the total cross section of transversely polarized vector mesons on 
nucleons.
\item  In the incoherent photoproduction $\gamma A\to \omega A^\prime$ ($A^\prime$ stands for the target excitation or its break-up products)  
the cross section on the nucleus is  the  sum of the photoproduction cross sections on  nucleons.  As a result $\omega$ mesons with both 
polarizations can be  produced\footnote{In this paper we consider the helicity reference frame.}. This can be used to study the interaction of longitudinally polarized vector mesons with matter~\cite{gevorkyan72}. 
\end{itemize}

The coherent and incoherent photoproduction of $\omega$ mesons will be described in Section 3 and Section 4.

\section{Coherent photoproduction}  
Coherent photoproduction of vector mesons on nuclei targets \ba \gamma + A\to V +A \ea 
has been studied for many years and is well described by Glauber multiple scattering theory~\cite{yennie78}.
The invariant momentum  transfer  in the  process (1)  can be expressed through the minimum 
longitudinal\footnote{The inverse of the longitudinal momentum transfer is called the coherence length   
$l_c=\frac{1}{q_L}$.}   momentum  $q_L$  and  the two dimensional transverse momentum  $\vec q_\perp$ defined as:
%-----------------------------
\ba t= (k - p)^2 & \simeq &  -(\frac{m_{\omega}^2}{2k})^2-4| \vec k |\; | \vec p |\; \sin^2{\frac{\theta}{2}}\nn
q_L^2 &= &(\frac{m_{\omega}^2}{2k})^2\nn
{\vec q}_\perp{}^2 & = & 4| \vec k |\; | \vec p |\; \sin^2{\frac{\theta}{2}},\ea
%-----------------------------
where $k$ and $p$ are the momenta of the beam photon and the vector meson, respectively. 
For the coherent reaction the nuclear target remains in the ground state after the meson is produced. The  production amplitude  of the vector meson with  helicity  $\lambda=0,\pm 1$ 
on the nucleus can be presented as
%-----------------------------
\ba F^{\lambda}(q_\perp{},q_L)&=&f_N^{\lambda}(0)\;F^{\lambda}_A(q_\perp,q_L)\nn
F^{\lambda}_A( q_\perp,q_L)&=&\int d^2b\;dz\; e^{i(q_Lz + \vec q_\bot\vec b)}\;\rho(b,z)\;{\rm exp}{ \{-\frac{\beta_\lambda}{2}\int_z^{\infty} dz^\prime\; \rho(b,z^\prime) \} }\ea
%-----------------------------
Here $f_N^{\lambda}(0)$ is the diffractive part of the photoproduction amplitude of the vector meson on the nucleon in the forward direction ($\theta = 0$),  $F^{\lambda}_A(q_\perp,q_L)$ 
is the nuclear form factor\footnote{The nucleon  density $\rho(b,z)$ is normalized to the  atomic weight $\int \rho(b,z)d^2bdz=A$.}  modified by the meson  absorption, and  $\vec b$ 
is the impact parameter. The complex parameter $\beta_\lambda = \sigma_{\lambda}(1-i\alpha_{\lambda})$ is related to the total meson-nucleon cross section $\sigma_{\lambda}$ 
and the ratio of the real to imaginary parts of the  forward meson-nucleon scattering amplitude  $\alpha_\lambda$. 

The coherent amplitude represents a sum of photoproduction amplitudes on individual nucleons. For isoscalar nuclei, the contribution to the coherent process from pion exchange can be 
neglected because the interaction amplitudes of a particle with isotopic spin one with protons and neutrons have the opposite signs and cancel out\footnote{For nuclei with unequal numbers 
of protons and neutrons, small corrections can be taken into account.}. The coherent production is dominated by the Pomeron exchange mechanism, where the photon helicity is preserved at 
small momenta transfer (s-channel helicity conservation), i.e., transverse photons ($\lambda=\pm 1$ in the helicity frame) produce  only transversely polarized vector mesons. The 
differential cross section of the coherent process can be written as follows:
%-----------------------------
\ba
\rho^A_{\lambda\lambda\prime}\frac{d\sigma_A}{dt}=\rho_{\lambda\lambda\prime} f_N^{*\lambda}(0) f_N^{\lambda\prime}(0) F_A^{*\lambda} (q_\perp,q_L)F_A^{\lambda\prime}(q_\perp,q_L), 
\ea
%-----------------------------
where $\rho^A_{\lambda\lambda\prime}, \rho_{\lambda\lambda\prime}$ are vector meson  spin density matrix elements for production on  nuclei and nucleons, respectively.
Using the relation between the diagonal elements of the spin density matrix  $\rho_{00}+\rho_{11}+\rho_{-1-1}=1$  we obtain  the well known expression ~\cite{yennie78} 
for the coherent photoproduction of vector mesons:
%-----------------------------
\ba
\frac{d\sigma_A}{dt} = \;\mid F_A( q_\perp,q_L,\sigma_T)\mid^2\; \left.\frac{d\sigma_N}{dt}\right|_{t=0}
\ea
%-----------------------------
Study of the coherent photoproduction of $\omega$ mesons  allows one to obtain the cross section of interactions of 
transversely polarized $\omega$ mesons with nucleons\footnote{ The authors of  the work~\cite{braccini70} used an angular 
distribution  $1+cos^2\theta$ for the decay mode $\omega\to \pi^0\gamma$ for both coherent and incoherent photoproduction, 
which is, strictly speaking, correct only for the coherent photoproduction.} $\sigma_T$ and to determine the $\omega$ photoproduction 
cross section on nucleons at zero angle for natural parity exchange, $\left.\frac{d\sigma_N}{dt}\right|_{t=0}$, which 
can be expressed in the vector dominance model as 
%-----------------------------
\ba
\left. \frac{d\sigma_N}{dt}\right |_{t=0}=\frac{4\pi}{\gamma_{\omega}^2}\frac{\alpha}{64\pi}\sigma^2_{\omega}(1+\alpha_{\omega}^2),
\ea
%-----------------------------
where  $\frac{\gamma_{\omega}^2}{4\pi}$ is the $\omega$ - photon coupling constant, $\alpha$ is the fine structure constant, $\sigma_{\omega}$ is the total 
$\sigma(\omega N)$ cross section, and $\alpha_{\omega}$ is the ratio of the real to imaginary parts of the $\omega N \to \omega N$ amplitude. 

The cross section $\left.\frac{d\sigma_N}{dt}\right|_{t=0}$ can be independently measured in $\omega$ photoproduction on nucleons using linearly polarized photons~\cite{ballam73}. 
The photon polarization allows one to distinguish contributions from the natural and unnatural parity exchange production mechanisms.
The beam of polarized photons used by the GlueX experiment will provide an opportunity to measure photoproduction cross sections of $\omega$ mesons on both 
nuclei and nucleons and therefore should help to sort out the contradictions in the measurements of the photon - omega coupling 
constant~\cite{sibirtsev03}.

\section{Incoherent photoproduction}
We consider photoproduction of $\omega$ mesons with different polarizations on nuclei in the reaction 
%-----------------------------
\ba
\gamma+A\to \omega +A'
\ea
%-----------------------------
where $A'$ denotes the nuclear target  excitation or the target break-up products. The incoherent photoproduction can be measured in the typical momentum transfer 
range $0.1\;{\rm GeV^2} < |t| < 0.5\;{\rm GeV^2}$, where the Glauber multiple scattering theory can be applied\footnote{At smaller momenta one has to take into account 
the suppression due to the exclusion principle (see e.g. ~\cite{gevorkyan12}), while  for a larger momentum transfer it is necessary to consider  incoherent  multiple 
scattering of  $\omega$ mesons.}. Two approaches based on the Glauber multiple scattering theory can be used to describe incoherent photoproduction.

$\bullet$ The first  model has been known for  many years~\cite{engel64} and was widely used~\cite{braccini70,margolis68,sibirtsev02}. In this model, $\omega$ mesons are 
produced on a nucleon with the momentum transfer $q$ and subsequently interact with the nuclear medium. We generalized this approach to account for potentially different 
absorptions of transversely and longitudinally polarized mesons. The cross section of the process (7) can be written as:
%-----------------------------
\ba
\frac{d\sigma_A(q)}{dt}&=&\frac{d\sigma_0(q)}{dt} (\rho_{00}N(\sigma_L)+(1-\rho_{00})N(\sigma_T))\nn
N(\sigma)&=&\int\frac{1-\exp(-\sigma\int{\rho(b,z)dz})}{\sigma}d^2b,\nn
\ea  
%-----------------------------
where  $d\sigma_0(q)/dt$ is the differential cross section of the  $\omega$ meson  photoproduction on nucleon, $\sigma_{T,L}$ is the total $\omega$-nucleon cross section for
longitudinally and transversely polarized mesons, and $\rho_{00}$ is the $\omega$ meson spin  density matrix element corresponding to the fraction of longitudinally polarized 
$\omega$ mesons. If $\sigma_T=\sigma_L$ the nuclear transparency has the well known form  $A_{\rm eff}=\frac{d\sigma_A}{dt}/{\frac{d\sigma_0(q)}{dt}}=N(\sigma)$.  Spin density 
matrix elements for photoproduction on nuclei $\rho^A_{00}$ and nucleons $\rho_{00}$ are related as
%-----------------------------
\ba
\rho_{00}^A=\frac{N(\sigma_L)}{\rho_{00}N(\sigma_L)+(1-\rho_{00})N(\sigma_T)}\rho_{00}
\ea
%-----------------------------
For  $\sigma_T=\sigma_L$,  $\rho_{00}^A = \rho_{00}$. For  $\sigma_T\neq\sigma_L$,  $\rho_{00}^A$ depends on the nuclear mass number A.

$\bullet$ Another approach to characterize incoherent photoproduction takes into account the interference of two amplitudes in the photoproduction process: production of a vector 
meson on one of the nucleons in the nucleus and production of the vector meson on the nucleon in the forward direction with subsequent scattering of the meson on another nucleon 
acquiring  transverse momentum~\cite{tarasov76}. A similar interference effect takes place in incoherent electroproduction of vector mesons and has to be taken into account in the 
studies of color transparency, i.e., weakening of vector meson absorption in nuclei with the increase of the  mass of the virtual photon $Q^2$. Studies of the color transparency 
are complicated by the dependence of the  incoherent cross section on the energy $\nu$ via the coherence length $l_c=\frac{1}{q_L}=\frac{2\nu}{Q^2+m_V^2}$,  leading  to a decrease 
of the incoherent  cross section at higher energies~\cite{hermes03,bob02,bob07}. Taking $\omega$  meson polarization into account, the incoherent cross section on nuclei is given 
by the expression:
%-----------------------------
\ba
\rho^A_{\lambda\lambda^\prime}\frac{d\sigma_A(q)}{dt}&=&\int d^2b \; dz \; {\rho(b,z)\phi^{*\lambda}(b,z)\phi^{\lambda^\prime}(b,z)}\nn
\phi^{\lambda}(b,z) &=&f^{\lambda}_p(q) \exp \{ -\frac{\beta_{\lambda}}{2}\int_z^{\infty} dz^\prime {\rho(b,z^\prime)} \}\nn
&-&\frac{2\pi}{ik}f^{\lambda}_p(0)f_s(q)
\int dz^\prime\; {\rho(b,z^\prime) e^{iq_L(z^\prime-z)}\exp \{-\frac{\beta_{\lambda}}{2}\int_{z^\prime}^{\infty}dz^{\prime\prime} \;{\rho(b,z^{\prime\prime})}\},}\nn
\ea  
%-----------------------------
where  $f_p^\lambda(q)$ and  $f_s^\lambda(q)$ are the amplitudes of the $\omega$ meson  photoproduction on a nucleon ($\gamma N\to \omega N$) and the elastic scattering 
($\omega N\to \omega N$), respectively. Assuming the same slopes  for the elementary  amplitudes  $f_p(q)$ and  $f_s(q)$ the incoherent cross section for the diagonal elements 
($\lambda=\lambda^\prime=0,\pm 1$) of the spin density matrix can be written as:
%-----------------------------
\ba
\rho^A_{\lambda\lambda}\frac{d\sigma_A(q)}{dt}&=&\rho_{\lambda\lambda}\frac{d\sigma_0(q)}{dt}\int d^2b\;dz\; {\rho(b,z)\mid\phi^{\lambda}(b,z)\mid^2}\nn
\phi^0(b,z)&=& \exp \{-\frac{\sigma_L}{2}\int_z^{\infty}dz^\prime \; {\rho(b,z^\prime) \} }\nn
\phi^{\pm 1}(b,z)&=&\exp \{-\frac{\sigma_T}{2}\int_z^{\infty}dz^\prime \; {\rho(b,z^\prime)} \}\nn
&-& \frac{\sigma_T}{2}\int^z dz^{\prime}\; {\rho(b,z^{\prime}) e^{iq_L(z^{\prime}-z)}\exp \{-\frac{\sigma_T}{2}\int_{z^{\prime}}dz^{\prime\prime} \; {\rho(b,z^{\prime\prime})}}\} \nn
\ea  
%-----------------------------
As can be seen the cross section for longitudinally polarized mesons does not depend on energy ($\phi^0$ does not depend on $q_L$).  There is no energy-dependent contribution to the 
production cross section from the amplitude interference effect because the  photoproduction amplitude of longitudinally polarized mesons at zero angle is zero. The interference is 
present in the photoproduction of transversely polarized $\omega$ mesons. Similar to equations (8) and (9) the incoherent cross section and the spin density matrix element $\rho_{00}^A$ 
for longitudinally polarized $\omega$ mesons can be written as
%-----------------------------
\ba
\frac{d\sigma_A(q)}{dt}&=&\frac{d\sigma_0(q)}{dt}\left (\rho_{00}N(\sigma_L)+(1-\rho_{00})W(q_L,\sigma)\right )\nn
W(q_L,\sigma)&=&\int {\rho(b,z)\mid\phi^{\pm 1}(b,z)\mid^2d^2bdz}\nn
\rho_{00}^A&=&\frac{N(\sigma_L)}{\rho_{00}N(\sigma_L)+(1-\rho_{00})W(q_L,\sigma_T)}\rho_{00}
\ea
%-----------------------------
In the limit of small photon energies, the term $W(q_L,\sigma_T)$ approaches $N(\sigma_T)$ from Eq.(8).

\begin{figure}[t]
\begin{center}
\includegraphics[width=1.0\linewidth,angle=0]{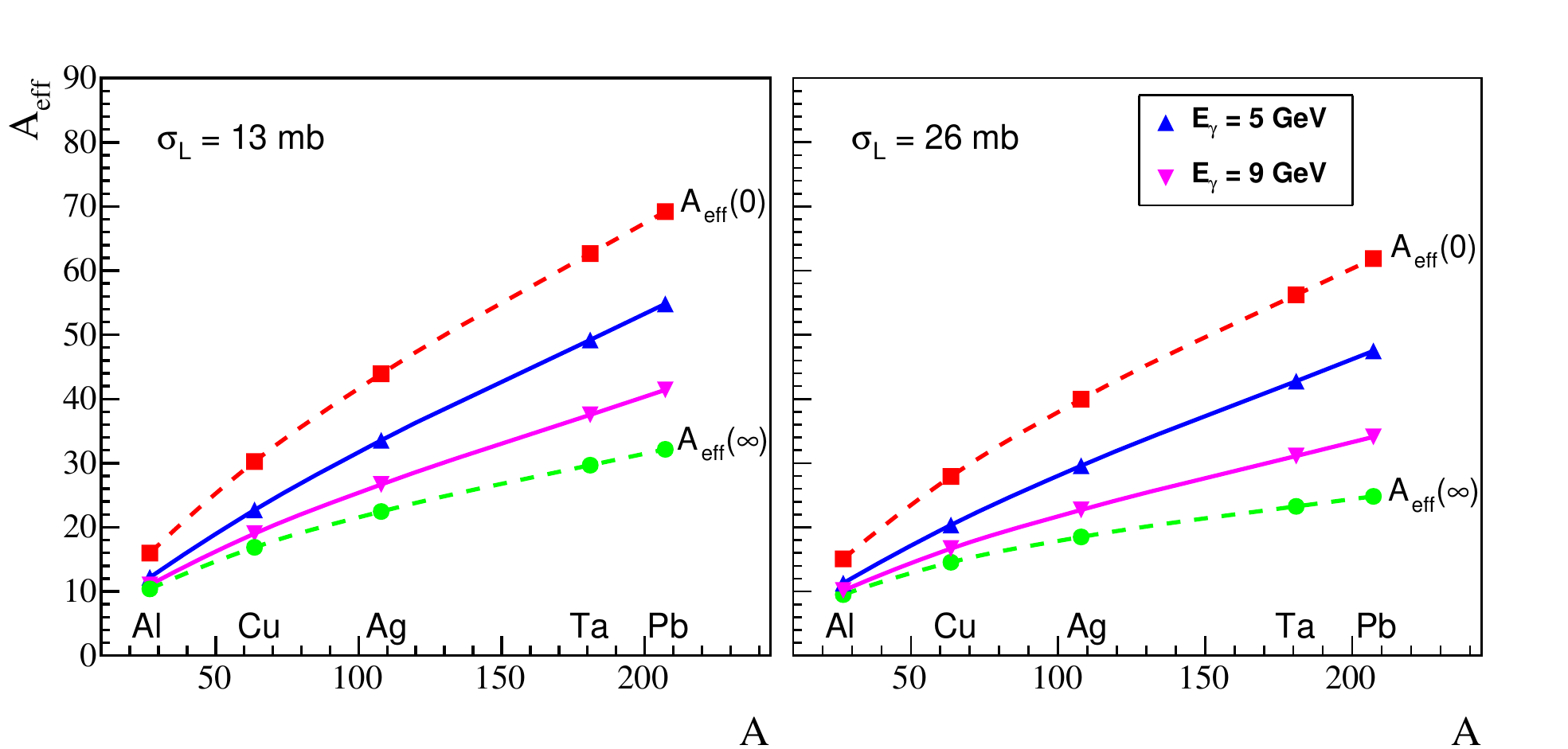}
\end{center}
\caption{Dependence of the nuclear transparency $A_{\rm eff}$  on the  mass number for $\sigma_L = {\rm 13\;{\rm mb}}$ (left) and  $\sigma_L = {\rm 26\;{\rm mb}}$ (right).
$A_{\rm eff}$ is shown for different photon energies: $A_{\rm eff}(\infty)$ and $A_{\rm eff}(0)$ correspond to infinite energy and energy-independent transparency, respectively.} 
\label{fig:aeff}
\end{figure}

\begin{figure}[t]
 \begin{center}
 \includegraphics[width=1.0\linewidth,angle=0]{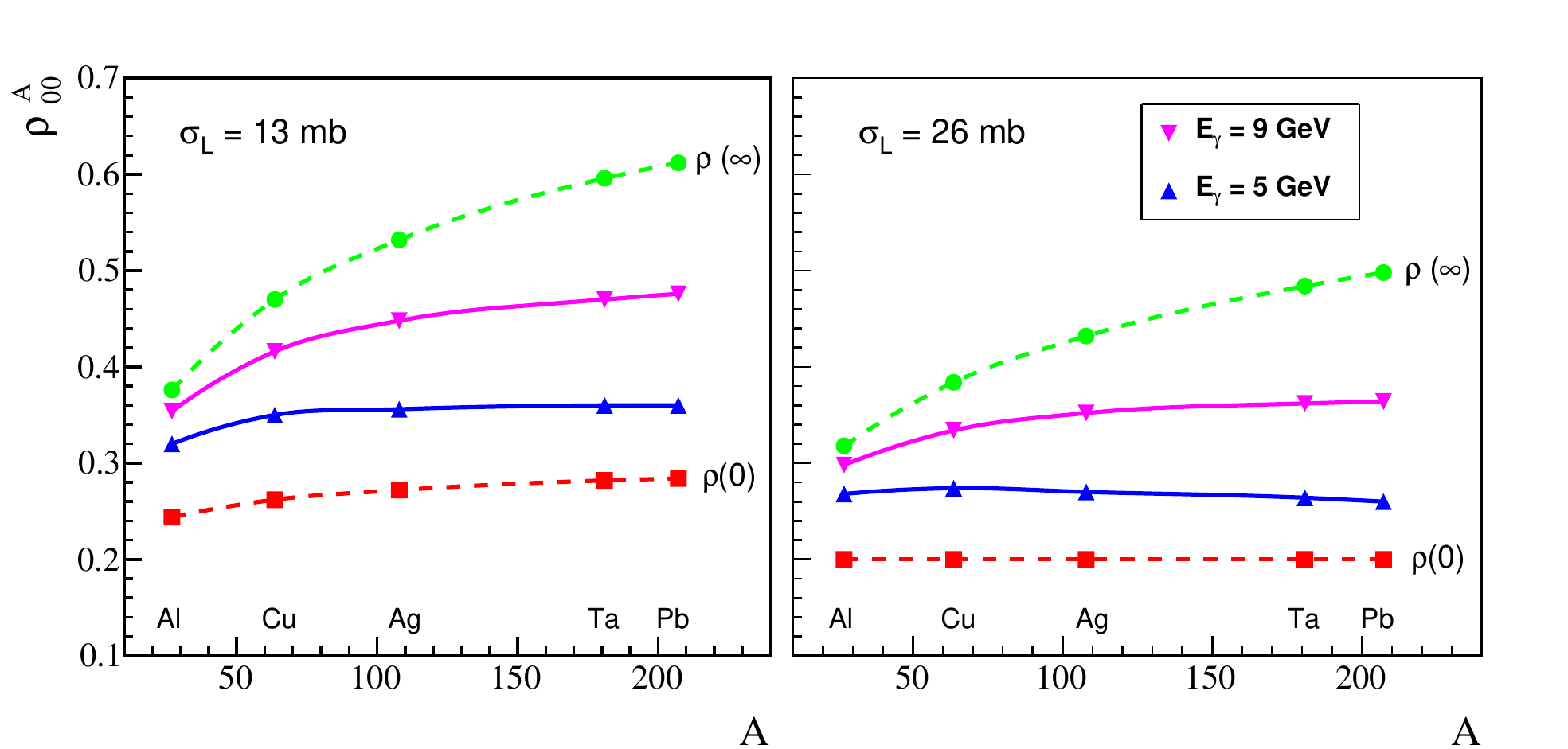}
 \end{center}
 \caption{A-dependence of the spin  density matrix element $\rho^{A}_{00}$ for  $\sigma_L = {\rm 13\;{\rm mb}}$ (left) and  $\sigma_L = {\rm 26\;{\rm mb}}$ (right). 
 $\rho^{A}_{00}$  is shown for different photon energies:  $\rho(\infty)$ and $\rho(0)$ correspond to infinite energy and energy-independent $\rho^{A}_{00}$, respectively.}
 \label{fig:rho}
 \end{figure}

\begin{figure}[t]
\begin{center}
\includegraphics[width=1.0\linewidth,angle=0]{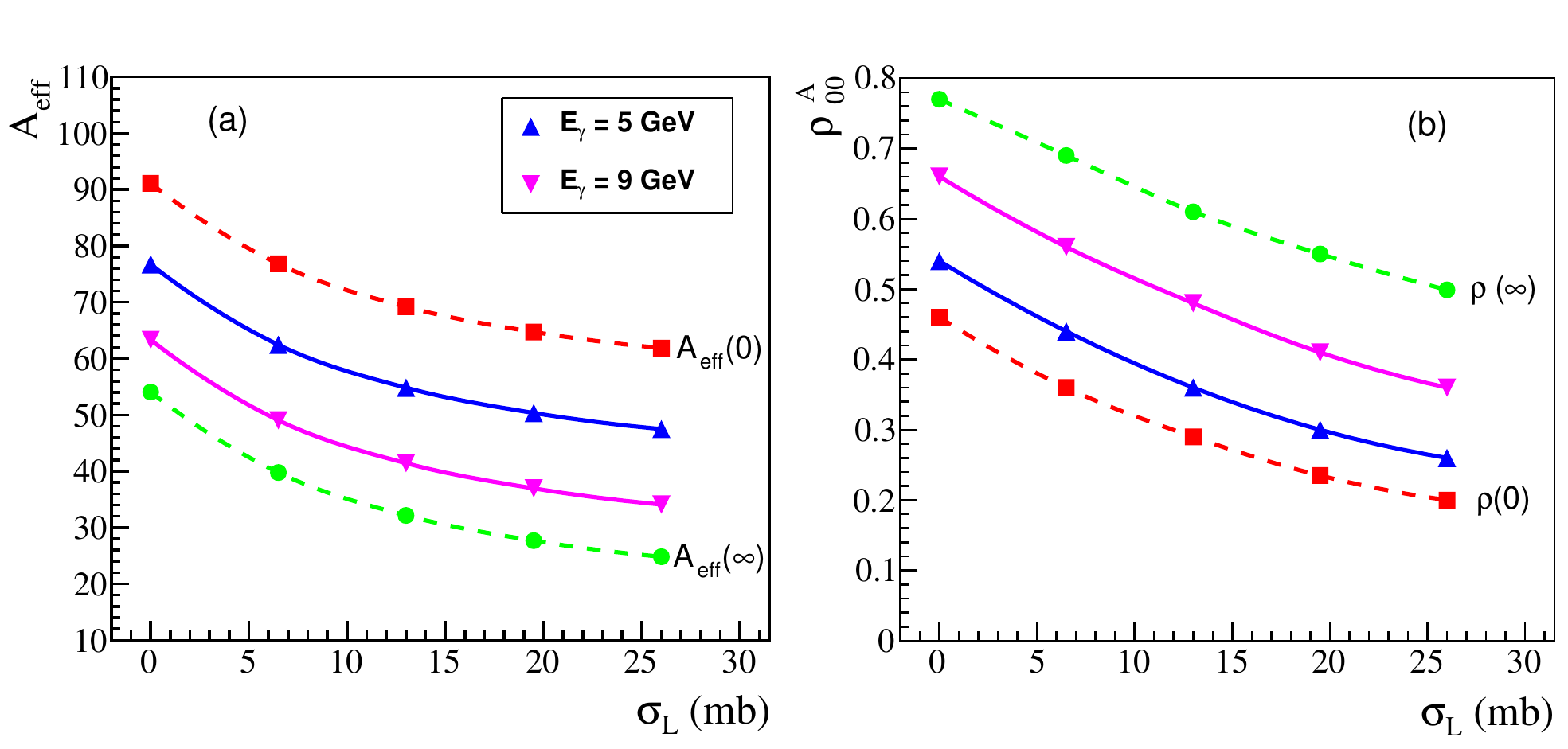}
\end{center}
\caption{ The nuclear transparency $A_{\rm eff}$ (a) and the spin  density matrix element  $\rho^{A}_{00}$ (b) as a function of $\sigma_L$ for lead nucleus.}
\label{fig:a_rho}
\end{figure}

The nuclear transparency $A_{\rm eff}=\frac{d\sigma_A}{dt}/\frac{d\sigma_0(q)}{dt}$ as a function of the mass number is presented in Fig.~\ref{fig:aeff} for $\sigma_L$ = 13 mb and 
$\sigma_L$ = 26 mb. The nuclear transparency is shown for two photon beam energies of 5 GeV and 9 GeV. Two boundary  conditions denoted as $A_{\rm eff}(\infty)$ and $A_{\rm eff}(0)$ 
correspond to infinite beam energy and the energy-independent  nuclear transparency given by Eq. 8. A-dependence of the density matrix element $\rho_{00}^A$ on nuclei for 
$\sigma_L$ = 13 mb and $\sigma_L$ = 26 mb and various beam energies is shown in Fig.~\ref{fig:rho}. The boundary conditions are denoted as $\rho(\infty)$ and $\rho(0)$.
Fig.~\ref{fig:a_rho} presents the nuclear transparency $A_{\rm eff}$ and the density matrix element $\rho_{00}^A$ as a function of the $\sigma_L$ for the lead nucleus target.

The value of the transverse $\omega$-nucleon cross section in these plots is set to $\sigma_T(\omega N) = 26\;{\rm mb}$ according to the measurements of  coherent photoproduction~\cite{braccini70}. 
We used the spin  density matrix element $\rho_{00}=0.2$ in the helicity frame as  measured in photoproduction on nucleons~\cite{ballam73}. For the  nuclear  density   we adopt 
the Woods-Saxon parametrization:
%-----------------------------
\ba
\rho(r)=\rho_0\frac{1}{1+\exp(\frac{r-R}{c})},
\ea 
%-----------------------------
with  $R=1.12A^{1/3}$ and c=0.545 fm.

\section{Summary}  
We discussed the motivation for $\omega$ meson photoproduction measurements on nucleons and different nuclei in the energy range $5\;{\rm GeV} < E_\gamma < 12\;{\rm GeV}$  
in the experimental Hall D at at Jefferson Lab. Coherent photoproduction of $\omega$ mesons on nuclei allows one to extract the total  cross section of 
the interaction of transversely polarized vector mesons with nucleons $\sigma_T (\omega N)$ and to measure the vector dominance model  $\omega$-photon coupling constant. The coupling constant can 
be independently obtained by measuring  photoproduction of $\omega$ mesons on nucleons using the Hall D beam of linearly polarized photons. These measurements 
should help to resolve certain inconsistencies in the results of the previous experiments and the SU(3) symmetry predictions~\cite{behrend70,braccini70,brodbeck78}.

Measurements of the photoproduction cross section and omega meson spin density matrix elements in the incoherent region ($|t| \geq 0.1\;{\rm GeV} $)  will allow to determine 
the total cross section of longitudinally polarized vector mesons with nucleons $\sigma_L (\omega N)$ and thus shed light on the impact of vector meson polarization on strong 
interactions. Availability of such measurements at different beam energies is essential to check the models of incoherent photoproduction. Neither the absorption of longitudinally 
polarized mesons, nor the spin density matrix elements on nuclei in photoproduction have been measured so far.

\section{Acknowledgments}

This work was supported by the Department of Energy. Jefferson Science Associates, LLC operated Thomas Jefferson National Accelerator Facility for the United 
States Department of Energy under contract DE-AC05-06OR23177. We would like to thank Stan Brodsky for helpful comments and suggestions. Sergey Gevorkyan appreciates 
the hospitality of Jefferson Lab  where this work has been done.

\clearpage
% ----------------------------------------------------------------------
% ----------------------------------------------------------------------
% ----------------------------------------------------------------------


\begin{thebibliography} {99}
\bibitem{yennie78} T.Bauer,R.Spital,D.Yennie,F.Pipkin, Rev.of Mod.Phys.50,261 (1978). 
\bibitem{ioffe00} B.L.Ioffe, A. G. Oganesian, Phys. Rev. D63, 096006 (2001).
\bibitem{ioffe02} B.L. Ioffe, arXiv:hep-ph/0209254 (2002).
\bibitem{forshaw10} J.R. Forshaw, R. Sandapen, JHEP 11,037 (2010).
\bibitem{forshaw12} J.R. Forshaw, R. Sandapen, Phys. Rev. Lett. 109,081601 (2012). 
\bibitem{brodsky09} G.F. de Teramond and S.J. Brodsky, Phys. Rev. Lett.102, 081601 (2009).
\bibitem{gevorkyan15} S. R. Gevorkyan, in preparation. 
\bibitem{glauber69} V. Franco, R. Glauber, Phys. Rev. Lett. 22, 370 (1969).
\bibitem{azhg10} L.Azhgirey et al. Part. Nucl.,Letters, 7,49 (2010).
\bibitem{strikman98} L. Gerland et al., Phys. Rev. Lett. 81, 762 (1998).
\bibitem{huefner00} J.Huefner, B.Z.Kopeliovich, A. V. Tarasov, Phys. Rev. D62, 094022 (2000).
\bibitem{jaus91} W. Jaus, Phys. Rev. D44, 2851 (1991).
\bibitem{leksin78} A.Arefyev et al.,Yad.Phys.27,161 (1978).
\bibitem{tarasov75} A. Pak, A. V.  Tarasov, Yad. Phys.22,91 (1975).
\bibitem{gevorkyan72} S. R. Gevorkyan, A.V. Tarasov, JETP Letters,15 684 (1972).
\bibitem{behrend70} H.Behrend et al. Phys.Rev.Lett.24,1246 (1970).
\bibitem{abramson78}J.Abramson et al. Phys.Rev.Lett.36,1428 (1976).
\bibitem{braccini70} P.Braccini et al. Nucl.Phys. B24, 173 (1970).
\bibitem{brodbeck78} T. J. Brodbeck et al., Nucl. Phys. B136,95 (1978).
\bibitem{elsa05} D.Trnka et al., Phys. Rev. Lett. 94,192303 (2005).
\bibitem{jlab10}M. H. Wood et al., Phys.Rev.Lett. 105,112301 (2010).
\bibitem{mosel10} S. Leupold, V. Metag, U. Mosel, Int. J. Mod. Phys. E. 19 147 (2010).
\bibitem{gluex}  JLab Experiment E12-06-102, (2006) \url{http://www.jlab.org/exp_prog/proposals/06/PR12-06-102.pdf.}
\bibitem{wolf69} K. Schilling, P. Seyboth, G. Wolf, Nucl. Phys. B15,397(1970).
\bibitem{ballam73} J. Ballam et al., Phys. Rev. D7,3150 (1973). 
\bibitem{sibirtsev03} A. Sibirtsev et al., Phys. Rev. C 67, 055201 (2003).
\bibitem{engel64} C.A. Engelbrecht, Phys. Rev. 133,B988 (1964).
\bibitem{tarasov76} A. V. Tarasov,  Phys. of Particles and Nuclei 7,771 (1976).
\bibitem{margolis68} K.S. Kolbig, B. Margolis, Nucl. Phys. B6, 85 (1968).
\bibitem{sibirtsev02} A. Sibirtsev, Ch. Elster, J. Speth arXiv:nucl-th/0203044 (2002).
\bibitem{gevorkyan12} S. R.  Gevorkyan et al., Phys.of Particles and Nuclei Letters 9, 18 (2012).
\bibitem{cornel69} G. McClellan et al. Phys. Rev. Lett. 23,554 (1969).
\bibitem{hermes03} A. Airapetian et al., Phys. Rev. Lett. 90, 052501 (2003).
\bibitem{bob02} B. Z. Kopeliovich, J. Nemchik, A. Schafer,A.Tarasov  Phys. Rev. C65, 035201 (2002).
\bibitem{bob07} B. Z. Kopeliovich, J. Nemchik, I. Schmidt,  Phys. Rev. C76, 015205 (2007).
\end{thebibliography}
\end{document}